\documentstyle [prl,aps,preprint]{revtex}

\begin{document}
\draft 
\title{Remarks on identical particles in 
de Broglie--Bohm theory}
\author{Harvey R.~Brown\footnote{Sub-Faculty of Philosophy, Oxford University, 
10 Merton Street, Oxford OX1 4JJ, U.K. (e-mail: 
harvey.brown@philosophy.oxford.ac.uk).}, 
Erik Sj\"{o}qvist\footnote{Department of Quantum Chemistry, Uppsala University,
Box 518, S--75120 Uppsala, Sweden (e-mail: 
eriks@Kvac.UU.SE).} and 
Guido Bacciagaluppi\footnote{Sub-Faculty of Philosophy, Oxford University, 
10 Merton Street, Oxford OX1 4JJ, U.K. (e-mail: 
guido.bacciagaluppi@philosophy.oxford.ac.uk).}} 
\maketitle
\begin{abstract}
It is argued that the topological approach to the (anti-)symmetrisation 
condition for the quantum state of a collection of identical particles, 
defined in the `reduced' configuration space, is particularly natural 
from the perspective of de Broglie--Bohm pilot-wave theory. 
\end{abstract}
\pacs{} 
\section{Introduction}
Probably no detailed treatment of identical particles in non-relativistic 
quantum mechanics has been more influential than that due to Messiah 
and Greenberg \cite{messiah64}. At least as far as the strict 
consequences of the indistinguishable nature of the particles are 
concerned, these authors are led more to clarify constraints on the 
observables associated with a collection of such particles than 
constraints on the state of the collection. In particular, unlike 
Girardeau \cite{girardeau65} and Mirman \cite{mirman73}, for example, 
they do not appear to regard it to be a consequence of the 
indistinguishability of the particles that the configuration space 
wave function satisfies the condition 
\begin{equation}
\psi ({\bf x}_{P^{-1}1}, {\bf x}_{P^{-1}2},...,{\bf x}_{P^{-1}N}) = 
e^{i\gamma} \psi ({\bf x}_{1}, {\bf x}_{2},...,{\bf x}_{N}) , 
\label{eq:symcond}
\end{equation}
where $N$ is the number of particles, $P$ is an arbitrary permutation 
on the set $\{ 1,2,...,N \}$, and $\gamma$ is a real number which 
may or may not depend on the point in the configuration space. 
This seems, indeed, to be a separate assumption; after all, the square of 
the modulus of the wave function in (\ref{eq:symcond}) is not 
(contrary to Ref. \cite{mirman73}, p.~113) equal to the probability 
of detecting $N$ particles in the spatial configuration ${\bf x}_{1}, 
{\bf x}_{2},...,{\bf x}_{N}$,\footnote{This probability is given by 
$\sum |\psi ({\bf x}_{P^{-1}1}, {\bf x}_{P^{-1}2},...,
{\bf x}_{P^{-1}N})|^{2}$, where the sum is over all $N!$ permutations 
$P$; see Ref. \cite{messiah62}, p. 584. This is of course also the
probability of finding any out of $N$ {\em distinguishable} particles
at the points ${\bf x}_1,{\bf x}_2, \ldots, {\bf x}_N$, respectively.} 
the expression of which should be invariant under a permutation of the 
particle labels. 

A separate and powerful approach to identical particles, however, 
does support the validity of (\ref{eq:symcond}), with the restriction
that $e^{i\gamma}$ be a global phase factor. This is the approach based 
on the use of a reduced configuration space, in which configurations
related by permutations are identified.\footnote{Notice that if the phase
$\gamma$ in (\ref{eq:symcond}) is global, that is independent of the point
in configuration space, then the wave function $\psi$ generally becomes
multi-valued (see below), unless of course $\gamma$ is an integer multiple 
of $\pi$.} As early as 
1927, Einstein \cite{einstein27} had voiced misgivings about 
Schr\"{o}dinger's use of the full configuration space formed 
by the $N$-fold Cartesian product of three-dimensional 
Euclidean space as the domain of the wave function, on the grounds 
that for a system of $N$ identical particles, there seemed to be a 
tension between considering configurations related by permutations 
as distinct and the recent results on particle statistics.
Note, however, that if wave functions are defined on the reduced 
configuration space, that would seem to force symmetry on the wave 
functions when written as functions on the full configuration space,
that is, $\gamma$ in (\ref{eq:symcond}) should be identically zero,
and only bosonic, not fermionic, statistics would be derivable. 

Nevertheless, Einstein was right. Later researchers were to vindicate
the use of the reduced configuration space in deriving the statistics
of identical particles. In particular, it was in the profound analysis 
of Laidlaw and DeWitt \cite{laidlaw71} and particularly Leinaas 
and Myrheim \cite{leinaas77} that the role of the non-trivial global 
topology (multiple connectedness) of the reduced configuration space 
was established. These insights are summarised in the 
next section. Suffice it to say here that in the topological 
approach the condition (\ref{eq:symcond}) above is valid, with
$e^{i\gamma}$ a global phase factor, even if it is not quite 
a simple consequence of the indistinguishability 
of the particles. And remarkably, it can be shown that if the 
physical space has at least three dimensions, then the phase 
factor in (\ref{eq:symcond}) must have the values $\pm 1$. In 
other words, the (anti-)symmetrisation condition on the wave 
function --- the origin of (fermionic) bosonic statistics --- is 
now seen to be related to the dimensionality of space, in contrast 
to the Messiah and Greenberg analysis wherein the 
(anti-)symmetrisation condition receives the status 
of a postulate. 

Some implications of this topological approach to the treatment 
of identical particles within the framework of the de Broglie--Bohm 
`pilot-wave' formulation of quantum theory 
\cite{debroglie27,bohm52,holland93,bohm93,cushing94,cushing96} 
have recently been studied \cite{sjoqvist95}. The purpose of 
the present paper is principally to stress one point not emphasised 
in \cite{sjoqvist95}, namely that the multiple connectedness of the 
reduced configuration space, which seems somewhat {\it ad hoc} 
in the standard formulation of the topological approach, receives 
a natural justification  within de Broglie--Bohm theory. 
Some brief, and hopefully pertinent remarks will also be made 
regarding the role of the full configuration space for 
distinguishable particles in this theory. (A separate investigation
of identical particles and their statistics from the point of view
of de Broglie--Bohm theory will be the subject of a further
publication.)
 
\section{Topological theory of identical particles}
Consider a physical system of $N$ identical particles that move 
in a $d$-dimensional Euclidean\footnote{For discussions of 
various non-Euclidean spaces in connection with the topological 
theory to be discussed below, see Refs \cite{einarsson90,li93}.} 
physical space $\Re^{d}$. (To avoid unnecessary 
complications we assume the wave function is a product of 
a spin part and a spatial part, and that the spins are parallel, 
as in the main argument of \cite{leinaas77}.) In standard quantum 
mechanics the wave function of the system is defined, as in 
the case of distinguishable particles, on the full product 
configuration space $\Re^{Nd} = \Re^{d} \times ... \times \Re^{d}$. 
Now the first occurrence of the claim that the (anti-)symmetrisation 
condition on the wave function associated with this system is 
related to the dimensionality $d$ appeared, to the best of our 
knowledge, in the work of Girardeau \cite{girardeau65}. This author 
assumed, as we have mentioned, the validity of (\ref{eq:symcond}) from 
the outset, which he regarded as a definitional property of 
identical particles. Having taken this step, Girardeau correctly 
allowed for the {\it a priori} possibility that the phase $\gamma$ 
depends on the configuration point, and exploited this possibility 
to construct a consistent, if somewhat idealised, model of three 
particles with hard cores moving in one spatial dimension with 
non-trivial boundary conditions, in which the wave function fails 
to satisfy the (anti-) symmetrisation condition. In an interesting 
argument, Girardeau further showed that such a failure is generally 
impossible when motion is extended to three dimensions. (The limitations 
of this proof will be seen shortly.)

An arguably more convincing argument to essentially the same end 
starts with Einstein's claim above, that the full configuration 
space $\Re^{Nd}$ contains, in the case of indistinguishable particles, 
redundant information. It would surely seem natural to consider instead 
the `reduced' space $\Re^{Nd} /S_{N}$, which is the quotient of 
$\Re^{Nd}$ obtained by the action of the symmetric group $S_{N}$ (the group 
of permutations $P$ above). Note that in taking $\Re^{Nd} /S_{N}$ 
as the appropriate configuration space for a quantum treatment of 
the system, one is effectively operating in an analogous fashion 
to the standard use of a reduced configuration (and hence phase) space 
in the classical solution to the so-called Gibbs paradox related to mixing 
of identical gases. In both cases, the identification of points 
related by 
a permutation of particle labels should perhaps not be considered 
as an inevitable consequence of the intrinsic nature of the particles, 
but rather as a reasonable step whose justification lies with 
the success of the theory built on it.\footnote{In a penetrating 
analysis of Gibbs' paradox, Hestenes \cite{hestenes70} has stressed 
that appeal to the reduced phase space for the purposes of 
determining entropy by counting states is ultimately justified 
only in relation to the physical operations of mixing and filtering  
of `like' gases. Whether the use of the reduced configuration 
space $\Re^{Nd} /S_{N}$ in the case of identical quantum particles 
carries more {\it a priori} justification than the analogous procedure 
in the case of classical gas particles is a moot point.} At any rate, 
the detailed procedure for applying quantum theory to $\Re^{Nd} /S_{N}$, 
rather than the full configuration space $\Re^{Nd}$, was demonstrated 
by Laidlaw and DeWitt \cite{laidlaw71} using Feynman formalism for 
$d=3$, and Leinaas and Myrheim \cite{leinaas77} using Schr\"{o}dinger 
quantisation for arbitrary $d$. 

Now an important first step in this theory is the removal from 
$\Re^{Nd} /S_{N}$ of all points corresponding to two or more 
particles occupying the same spatial position at the same instant. 
The removal of such coincidence points, which are singular in 
$\Re^{Nd} /S_{N}$, is sometimes justified by considering the
particles to be `impenetrable', but of course impenetrability
is not a direct consequence of 
indistinguishability (and does not seem to hold for bosons). For 
the moment we shall simply assume that the appropriate configuration 
space is indeed the multiply connected set $Q = \Re^{Nd} /S_{N} - 
\Delta$, where $\Delta$ is the set of coincidence 
points.\footnote{The insistence on removing the singular points 
appears to be part of standard wisdom \cite{wu84}. We just wish to 
note, however, that one can subdivide paths in different equivalence 
classes according to their winding numbers around the singularities
and corresponding distinct values of the phase factor $e^{i\gamma}$,
even retaining the singular points, if one sets the wave function zero
(as a necessary condition for particles other than bosons)
at the singular points, thus allowing for a discontinuous behaviour 
of the phase when paths are continuously deformed through a
singular point. (The paths will be considered equivalent if amplitudes 
and phases separately are continuous under deformation.)} 

In order to see the implications of the multiple connectedness
of $Q$, it is convenient (though by no means necessary) to use 
the Feynman path integral 
formalism. Feynman paths that connect two points $x'$ and $x$ 
in $Q$ cannot in general be continuously deformed into each 
other without crossing a point in $\Delta$. This means 
that paths in the Feynman propagator $K(x,t;x',t')$ divide 
into homotopy classes each of which consists of paths that 
are homotopically equivalent. Explicitly, denoting the homotopy 
classes by $[\alpha ]$ we may write \cite{laidlaw71,wu84,morandi92}
\begin{equation}
K(x,t;x',t') = 
\sum_{[\alpha ]} \chi ([\alpha ]) K_{[\alpha ]} (x,t;x',t') , 
\end{equation} 
with $\chi ([\alpha ])$ phase factors and 
$K_{[\alpha ]} (x,t;x',t')$ the Feynman propagator formed 
by the paths in $[\alpha ]$. It is precisely the possibility 
of having different $\chi ([\alpha ])$ for different 
$[\alpha ]$ that physically distinguishes quantum 
mechanics on a multiply connected space (such as $Q$) from 
quantum mechanics on a simply connected space (such as 
$\Re^{Nd}$). A well known example of quantum mechanics 
on a multiply connected space is the Aharonov--Bohm effect 
\cite{aharonov59} where the $\chi ([\alpha ])$ are given by 
$\exp (in_{[\alpha ]} \Phi )$, with $\Phi$ the magnetic flux 
and $n_{[\alpha ]}$ the winding number that characterises 
$[\alpha ]$ around the flux line, which is the singular point. 
In the case of identical particles, the possible values of 
$\chi ([\alpha ])$ are determined by the topology of $Q$. 
In three dimensions $(d=3)$, $Q$ can be shown to be doubly 
connected and therefore $\chi [\alpha ] = \pm 1$, whereas
in the two-dimensional case $(d=2)$, $Q$ is infinitely 
connected and any $\chi [\alpha ]$ value may occur. 

Translating back into the language of wave functions on 
the full configuration space $\Re^{Nd}$, the implication 
of the topological considerations above is that in three 
dimensions, the phase factor in (\ref{eq:symcond}) is 
necessarily $\pm 1$ (corresponding to bosonic and fermionic 
behaviour), but that in the case of particles constrained 
to move in two dimensions, the (global) phase factor 
could be arbitrary (and the wave function is 
multi-valued). (Equivalently, a `statistics field' is introduced 
to characterise the different types of identical particles,
and the wave function is always single-valued --- in fact 
symmetric.) Identical particles in $\Re^{2}$ that obey 
intermediate statistics between the bosonic and fermionic 
cases are called anyons \cite{wilczek82}. The theory of 
such particles has been successfully applied to the fractional 
quantum Hall effect (see Refs \cite{tsui82,laughlin83,arovas84}), 
and provides considerable justification for the topological 
approach outlined above. Anyons were independently predicted 
using a quite different approach in Ref.~\cite{goldin81}.
This approach is not dynamical in the sense of the de Broglie--Bohm
one sketched below, but it does lead directly to $Q$ as a possible
configuration space.

It may be of interest here briefly to compare these results 
with the analysis of Girardeau \cite{girardeau65} mentioned 
at the beginning of this section. Girardeau attempted to show 
that generally the phase factor in (\ref{eq:symcond}) equals 
$\pm 1$, and that it is only in the case of three dimensions 
$(d=3)$ that its value must be global, i.e. independent of the 
point in the full configuration space. In the case of particles 
constrained to move in two dimensions, this result is 
inconsistent with the existence of anyons, for which the phase 
factor in question is global, being a property of the kind of 
particles involved, but not equal to $\pm 1$. However, in 
his proof, Girardeau assumed the availability of real-valued 
wave functions, and it can be shown in the topological approach 
that systems of anyons must have complex-valued wave functions 
due to the broken time reversal symmetry associated with statistics 
phase factors different from $\pm 1$ (see Ref. \cite{morandi92}, 
pp. 134--135). 

\section{De Broglie--Bohm theory}
Let us return to the issue at the heart of the topological 
approach which is that of the removal of the set $\Delta$ of 
coincidence points from the reduced configuration space 
$\Re^{Nd} /S_{N}$, rendering $Q$ multiply connected. In 
the theory of anyons, it has been conjectured 
\cite{aitchison91,stachel97} that short-range repulsive 
forces are at work between the particles; the adoption of 
such a view in the general case of identical particles appears 
however {\it ad hoc}. (Note that within the topological approach 
it is only strictly necessary for non-bosonic behaviour.) We now 
argue that such {\it ad hoc}ness is removed in the framework of 
de Broglie--Bohm pilot-wave theory. 

Recall that the de Broglie--Bohm formulation of non-relativistic 
quantum mechanics 
\cite{debroglie27,bohm52,holland93,bohm93,cushing94,cushing96} 
posits, besides the wave function on $\Re^{Nd}$ for an isolated 
system of $N$ particles (in the case of such a system being in 
a pure state), a collection of $N$ point corpuscles. It is 
the role of the wave function (`pilot-wave'), itself a solution 
of the time dependent Schr\"{o}dinger equation, to determine 
the instantaneous velocities of the corpuscles through the 
guidance equations  
\begin{equation}
m_{k} \dot{{\bf X}}_{k} = \nabla_{{\bf x}_{k}} S |_{x=X} ; 
\, \, \, \, k=1,...,N , 
\label{eq:ge}
\end{equation}
where $\dot{{\bf X}}_{k}$ is the velocity of the $k$th corpuscle 
with mass $m_{k}$, $S=S(x,t)$ is the phase of the pilot-wave 
(in units of $\hbar$) at 
the configuration point $x=({\bf x}_{1},...,{\bf x}_{N})$ and 
$X=({\bf X}_{1},...,{\bf X}_{N})$ is the instantaneous 
configuration of the $N$ corpuscles. (Note that in the case 
of an external field acting on the system and which is 
represented by a vector potential, the right-hand side 
of (\ref{eq:ge}) will incorporate an additive term linear 
in the vector potential, which in particular makes 
(\ref{eq:ge}) gauge-invariant.)

A feature of pilot-wave theory that is important for our 
purposes is this. Whereas in standard quantum mechanics the 
points of the configuration space are interpreted as the possible 
results of position measurements or detections of the $N$ particles, 
which are generally non-localised in $\Re^{d}$ prior to the 
measurements, in de Broglie--Bohm theory they are normally 
interpreted as the possible positions of the $N$ de Broglie--Bohm 
corpuscles. That is to say, they are truly objective configurations 
(as in classical theory), and make no reference to measurement 
or detection processes.  

In a recent analysis \cite{sjoqvist95} it has been demonstrated 
that the reduced configuration space approach is a natural 
framework for identical particles in de Broglie--Bohm 
theory. The identity of the particles implies that any two 
distinct initial configurations of the corpuscles that differ 
only by a permutation yield the same set of corpuscle 
trajectories ${\bf X}_{k} (t)$, $k=1,...,N$, in $\Re^{d}$. 
Again this shows that the full product configuration space 
$\Re^{Nd}$ contains redundant information. Indeed, taking 
the restriction $\Re^{Nd} / S_{N}$ as the configuration space, 
we obtain the same set of trajectories $\{ {\bf X}_{k} (t) \}$ 
in the physical space $\Re^{d}$. The physical configuration space 
for a system of $N$ identical particles in de Broglie--Bohm 
theory is therefore given by $\Re^{Nd} / S_{N}$. 

What is the status of the singular points with respect 
to the de Broglie--Bohm trajectories? Consider for simplicity 
the case of $N=2$, and suppose that the two corpuscles have 
distinct positions in $\Re^{d}$ at time $t_{0}$ and that 
they coincide at some finite time $t>t_{0}$. The single 
trajectory in $\Re^{2d} /S_{2}$ which contains the coincidence 
point at $t$ will generate two trajectories in the full 
configuration space, the point in one at any instant being 
obtained from that in the other `mirror' trajectory by a 
permutation of particle labels. But such trajectories in 
$\Re^{2d}$ will cross at the coincidence point, a possibility 
that is ruled out by the first order nature of the guidance 
equations (\ref{eq:ge}). It follows that the coincidence points 
are inaccessible from non-coincidence configurations at $t_{0}$. 
Conversely, the time-reversal symmetry of the de Broglie--Bohm 
trajectories implies that two identical corpuscles that start 
at the same point in space (as two bosonic corpuscles could)
will remain coincident forever.
Intuitively, since these initial conditions 
(pilot-wave and spatial position) are entirely symmetric with 
respect to the two corpuscles, and we are assuming that the 
symmetry of the pilot-wave is preserved over time, we do not 
expect their future or past trajectories to differ, and so the 
corpuscles coincide forever, if at all. 

If need be, we can put this argument on a more rigorous footing. 
Writing $\psi = Re^{iS/\hbar}$ for the system in centre of mass 
coordinates ${\bf x}={\bf x}_{1}-{\bf x}_{2}$ and 
${\bf x}_{CM}=({\bf x}_{1}+{\bf x}_{2})/2$, it follows from the 
symmetry or antisymmetry (or given the anyonic phase relation) 
of $\psi$ that $S(-{\bf x})=S({\bf x})+\gamma$, which implies 
that $\nabla_{{\bf x}} S({\bf x}) = 0$ at ${\bf x} = 0$. We then 
obtain immediately from the relation $\nabla_{{\bf x}} = 
(\nabla_{{\bf x}_{1}} -\nabla_{{\bf x}_{2}})/2$ that 
$\nabla_{{\bf x}} S = m (\dot{{\bf X}}_{1} - \dot{{\bf X}}_{2})/2 
= 0$ at ${\bf x} = 0$. From the vanishing relative velocity at 
the coincidence points and the first-order nature of the guidance
equation (\ref{eq:ge}) one concludes that two initially separated 
corpuscles for identical particles cannot reach the same point in 
the physical space at the same time

The generalisation of this conclusion to the case of arbitrary 
$N$ is straightforward. Thus, the sets corresponding to the 
$M$-point coincidences for any $M\leq N$, with their union $\Delta$,
as well as the set $Q = \Re^{Nd} /S_{N} - \Delta$, are invariant 
submanifolds of the reduced configuration space $\Re^{Nd} /S_{N}$
under the action of the de Broglie--Bohm dynamics (\ref{eq:ge}).
In particular, the space $Q$ of regular points can be used consistently 
as a configuration space for a de Broglie--Bohm theory. Further, removal 
of the sets of $M$-point coincidences seems physically well motivated,
since they correspond to motions for which $M$ particles coincide for 
all times --- which would appear as the motion of one particle of
$M$-fold mass and charge. (And removal of these sets does not affect
the statistical predictions of de Broglie--Bohm theory, since they
have total $|\psi|^2$-measure zero.) We thus argue that within the 
topological approach to 
identical particles the removal of the set $\Delta$ of 
coincidence points from the reduced configuration space 
$\Re^{Nd} /S_{N}$ thus follows naturally from de Broglie--Bohm 
dynamics as it is defined in the full space $\Re^{Nd}$. 

We finish this section by removing a possible source of confusion. 
It is well known that the nodal set (that is, the set of zeros) of 
the wavefunction can be shown to be accessible from the outside at
most for a set of initial conditions of $|\psi|^2$-measure zero
(see Ref.~\cite{berndl95}). It is also well known that the wave 
function of a system of bosons 
(unlike that of a system of fermions) need not vanish at a 
coincidence point in $\Re^{Nd}$. As a consequence it  
has been claimed that the de Broglie--Bohm trajectories of 
bosonic corpuscles may cross in the physical space (see Ref. 
\cite{holland93}, p. 284). But we have just seen how 
de Broglie--Bohm dynamics secures the inaccessibility of 
singular points in $\Re^{Nd} /S_{N}$, or coincidences in 
$\Re^{Nd}$, without qualifications and irrespective
of whether they also correspond to nodal points. So, if the
above claim is meant in the sense that particles starting from 
different points in space can cross, it is contradicted by our 
above result. If on the contrary it means merely that 
de Broglie--Bohm theory in the full configuration space is able
to describe coincident bosons, then it is obviously correct. 
We have only suggested that it would be more natural to exclude
such coincidences, because we have shown they would hold 
forever. The difference between the case of bosons and that of 
fermions lies not in a non-zero probability for coincidence of 
bosons (given that the set of coincidences has $|\psi|^2$-measure 
zero irrespective of whether or not the wave function vanishes 
at the coincidences); it lies rather in
the fact that while at a node the phase of the wave function is
ill-defined, and thus the de Broglie--Bohm dynamics breaks down
for coincident fermions, the trajectories of coincident bosons, if 
one should wish to retain them, would be well-defined for all times.

\section{Distinguishable particles} 
One can easily convince oneself of the fact that the first-order 
nature of the guidance equations (\ref{eq:ge}) implies that the 
guidance equations will continue to be first order when the wave 
function of the system of identical particles is defined on 
$\Re^{Nd} /S_{N}$, so that for any point in $\Re^{Nd} /S_{N}$ 
there will be a single curve in this space containing it. This 
orbit gives rise to $N!$ curves in $\Re^{Nd}$, each related to 
any other by a permutation of particles labels. But suppose the 
particles are distinguishable; then the guidance equations tell 
us that curves in $\Re^{Nd}$ that at a given instant contain points 
related by such permutations generally do not continue to be thus
related at other times, and so cannot be generated from a 
curve in $\Re^{Nd} /S_{N}$. 

The fact that the correct quantum mechanical treatment of a system 
of $N$ distinguishable particles requires the wave function to be 
defined on $\Re^{Nd}$, rather than $\Re^{Nd} /S_{N}$, suggests 
strongly that the hypothetical corpuscles in de Broglie--Bohm 
theory associated with such a system are pairwise distinct, or 
`labelled'. After all, given both the meaning of the configuration 
space in this theory (see above) and the apparent success of the 
topological approach for identical particles, it would be awkward
(but perhaps not inconsistent --- see footnote 4) 
to maintain that the point corpuscles were intrinsically identical, 
apart from their spatial positions, while assuming that the correct 
domain of the pilot-wave for the system is $\Re^{Nd}$. 

What properties possessed by the de Broglie--Bohm corpuscles in 
this case serve to label them? The obvious answer seems to be 
that it is the same properties that distinguish the particles 
in the conventional theory, viz mass, charge, magnetic moment 
etc. Indeed what else could they be? But it has not escaped 
notice that certain interference experiments involving single 
particles seem to suggest that these dynamical properties pertain 
to the pilot-wave.\footnote{See Refs \cite{brown95,brown96}. As 
an example involving charge \cite{brown95}, consider the 
Aharonov--Bohm effect \cite{aharonov59} referred to in the previous 
section. The expression for the phase shift due to the flux in the 
shielded solenoid depends on charge being present on spatial loops 
within the support of the wave function and
enclosing the solenoid, whereas the trajectory of the de 
Broglie--Bohm corpuscle associated with the charged particle 
does not encircle the solenoid (see Ref. \cite{bohm93}, Sec. 3.8).} 
This does not mean that they cannot also belong to the corpuscle. 
Arguments in favour of this `principle of generosity' (related, 
e.g., to the `inertia' of the corpuscles) in de Broglie--Bohm theory
--- the assignment of such properties as mass and charge to both
the pilot-wave and the corpuscle --- have indeed been given
in the literature (for a review see \cite{brown96}). The dynamical 
considerations raised in the previous paragraphs provide in our 
opinion another argument in favour of the principle of generosity.  
\vskip 0.5 cm
Remarks made by Tim Maudlin during the 1995 conference 
`Quantum Theory Without Observers', held in Bielefeld, 
Germany, alerted one of us (H.R.B.) to the importance of 
considering the role of the configuration space in understanding 
the nature of the de Broglie--Bohm corpuscles, and were the 
inspiration for Section IV of this paper. We are also grateful to 
Jerry Goldin and Peter Holland for discussion and comments. 
Any errors are, of course, our own responsibility. E.S.\ acknowledges 
financial support from the Wenner--Gren Foundation and G.B.\ a generous
Postdoctoral Research Fellowship from the British Academy.


\begin{references}
\bibitem{messiah64} A.M.L. Messiah and O.W. Greenberg, 
Phys. Rev. 136 (1964) B 248. 
\bibitem{girardeau65} M.D. Girardeau, 
Phys. Rev. 139 (1965) B 500. 
\bibitem{mirman73} R. Mirman, 
Nuovo Cimento 18 B (1973) 110. 
\bibitem{messiah62} A.M.L. Messiah, 
Quantum Mechanics,  Vol. 2 (North-Holland, Amsterdam, 1962). 
\bibitem{einstein27} \'{E}lectrons et Photons, Rapports et 
discussions du cinqui\`{e}me Conseil de Physique Solvay 
(Gauthier-Villars, Paris, 1928) p. 256. 
Excerpt of discussion translated and reprinted in:  
Niels Bohr, Collected Works, Vol.~6, ed. J.~Kalckar
(North-Holland, Amsterdam, 1985) pp. 102--103.
\bibitem{laidlaw71} M.G.G. Laidlaw and C.M. DeWitt, 
Phys. Rev. D 3 (1971) 1375. 
\bibitem{leinaas77} J.M. Leinaas and J. Myrheim, 
Nuovo Cimento 37 B (1977) 1.  
\bibitem{debroglie27} L. de Broglie, J. de Phys. 8 (1927) 225.
\bibitem{bohm52} D. Bohm, Phys. Rev. 85 (1952) 166; 
Phys. Rev. 85 (1952) 180.
\bibitem{holland93} P.R. Holland, The Quantum Theory of Motion 
(Cambridge University Press, Cambridge, 1993).
\bibitem{bohm93} D. Bohm and B.J. Hiley, The Undivided Universe 
(Routledge, London, 1993). 
\bibitem{cushing94} J.T. Cushing, Quantum Mechanics: Historical 
Contingency and the Copenhagen Hegemony 
(The University of Chicago Press, Chicago, 1994). 
\bibitem{cushing96} Bohmian Mechanics and Quantum Theory: 
An Appraisal, edited by J.T. Cushing, A. Fine and S. Goldstein  
(Kluwer, Dordrecht, 1996). 
\bibitem{sjoqvist95} E. Sj\"{o}qvist and H. Carlsen, 
Phys. Lett. A 202 (1995) 160. 
\bibitem{einarsson90} T. Einarsson, 
Phys. Rev. Lett. 64 (1990) 1995. 
\bibitem{li93} D.-P. Li, 
Nucl. Phys. B 396 (1993) 411. 
\bibitem{hestenes70} D. Hestenes, 
Am. J. Phys. 38 (1970) 840. 
\bibitem{wu84} Y.-S. Wu, 
Phys. Rev. Lett. 52 (1984) 2103. 
\bibitem{morandi92} G. Morandi, 
The Role of Topology in Classical and Quantum Physics 
(Springer-Verlag, Berlin, 1992).
\bibitem{aharonov59} Y. Aharonov and D. Bohm, 
Phys. Rev. 115 (1959) 485. 
\bibitem{wilczek82} F. Wilczek, 
Phys. Rev. Lett. 49 (1982) 957. 
\bibitem{tsui82} D.C. Tsui, H.L.Stormer and A.C. Gossard, 
Phys. Rev. Lett. 48 (1982) 1559. 
\bibitem{laughlin83} R.B. Laughlin, 
Phys. Rev. Lett. 50 (1983) 1395. 
\bibitem{arovas84} D. Arovas, J.R. Schrieffer and F. Wilczek, 
Phys. Rev. Lett. 53 (1982) 722. 
\bibitem{goldin81} G.A. Goldin, R. Menikoff and D.H. Sharp,
J. Math. Phys. 22 (1981) 1664.
\bibitem{aitchison91} I.J.R. Aitchison and N.E. Mavromatos, 
Contemp. Phys. 32 (1991) 219. 
\bibitem{stachel97} J. Stachel, Feynman paths and quantum 
entanglement: is there any more to the mystery?, in: 
Potentiality, Entanglement and Passion-at-a-Distance, 
edited by R.S. Cohen, M. Horne and J. Stachel (Kluwer, 
Dordrecht, 1997). 
\bibitem{berndl95} K. Berndl, D. D\"{u}rr, S. Goldstein,
G. Peruzzi and N. Zangh\`{\i}, Comm. Math. Phys. 173 
(1995), 647.
\bibitem{brown95} H.R. Brown, C. Dewdney and G. Horton, 
Found. Phys. 25 (1995) 329.  
\bibitem{brown96} H.R. Brown, A. Elby and R. Weingard, Cause 
and effect in the pilot-wave interpretation of quantum mechanics, 
in: Ref. \cite{cushing96}. 
\end{references}
\end{document}